\documentclass[prb,twocolumn,showpacs,linenumbers,amsbsy,floatfix,amsbsy,floatfix,superscriptaddress]{revtex4}
\usepackage{epsfig,color}
\usepackage[cp1250]{inputenc}
\usepackage{amsmath}
\usepackage{amssymb}
\usepackage{color}

\begin{document}

\title{All electrically controlled quantum gates for single heavy hole spin qubits}

\author{P. Szumniak}
\email{pawel.szumniak@gmail.com}
\affiliation{AGH University of Science and Technology, Faculty of
Physics and Applied Computer Science,\\
al. Mickiewicza 30, 30-059 Krak\'ow, Poland}
\author{S. Bednarek}
\affiliation{AGH University of Science and Technology, Faculty of
Physics and Applied Computer Science,\\
al. Mickiewicza 30, 30-059 Krak\'ow, Poland}
\author{J. Pawłowski}
\affiliation{AGH University of Science and Technology, Faculty of
Physics and Applied Computer Science,\\
al. Mickiewicza 30, 30-059 Krak\'ow, Poland}
\author{B. Partoens}
\affiliation{Department of Physics, University of Antwerp, Groenenborgerlaan 171, 2020 Antwerp, Belgium}
\date{\today}

\begin{abstract}

   In this paper, several nanodevices which realize basic single heavy hole qubit operations are proposed and supported by time dependent self consistent Poisson-Schr\"{o}dinger calculations using a four band heavy hole-light hole model. In particular we propose a set of nanodevices which can act as Pauli $X$, $Y$, $Z$ quantum gates and as a gate that acts similar as a Hadamard gate (i.e. it creates a balanced superposition of basis states but with an additional phase factor) on the heavy hole spin qubit. We also present the design and simulation of a gated semiconductor nanodevice which can realize an arbitrary sequence of all these proposed single quantum logic gates.
   The proposed devices exploit the self-focusing effect of the hole wave function which allows for guiding the hole along a given path in the form of a stable soliton-like wave packet. Thanks to the presence of the Dresselhaus spin orbit coupling, the motion of the hole along a certain direction is equivalent to the application of an effective magnetic field which induces in turn a coherent rotation of the heavy hole spin.
   The hole motion and consequently the quantum logic operation is initialized only by weak static voltages applied to the electrodes which cover the nanodevice.
   The proposed gates allow for an all electric and ultrafast (tens of picoseconds) heavy hole spin manipulation and give the possibility to implement a scalable architecture of heavy hole spin qubits for quantum computation applications.

\end{abstract}


\maketitle

\section{INTRODUCTION}

   The idea to realize quantum computers has attracted an enormous attention and effort of theoreticians and experimentalists in the last years. Among many appealing proposals for the physical realization of quantum computation, solid state spin based implementations seem to be particularly interesting and promising\cite{qd2, qd2a}. The spin state of an electron which is confined in a semiconductor nanostructure like a quantum dot or a quantum wire is considered to be a perfect candidate as carrier of a quantum bit of information\cite{qd1}. The realization of many state of the art experiments where an electron spin qubit can be prepared in a certain spin state, stored, manipulated and read out\cite{qd3, qd4, qd4a,qd5, qd6, qd7, qf7a, qd8, qd29, 1006,1020} show the enormous progress that has been made in the field in the last decade.

   Very challenging demands for the physical realization of quantum computation\cite{qd1a} are to obtain long living qubits which are immune to decoherence and to develop control methods which allow for a high fidelity and ultrafast qubit manipulation. Furthermore, the scalability requirement of the physical implementation of quantum computation imposes that one has to be able to control each qubit in the quantum register in an individual, selective manner as well as to couple long distant qubits so that also two-qubit gates can be realized.

   The main difficulty related to the use of the electron spin as a qubit is its relatively short coherence time. In most quantum dot structures the spin of the confined electron experiences a contact hyperfine interaction with a large number of nonzero nuclear spins of the host material. This results in electron spin decoherence\cite{qd60,qd60a,qd60b,1014a} and if no special effort is made an electron spin qubit loses its coherence in nanoseconds.

Several appealing ideas have been proposed and successfully applied to overcome the fast electron spin decoherence process \cite{qd50} such as the application of spin echo techniques \cite{1013_0, 1013, 1013a} or the preparation of the nuclear spins of the host material in a special narrow state  \cite{1014,1014a, 1015, 1016, 1017}. A straightforward approach to avoid the interaction with nuclear spins is to confine the electron in a nuclear-spin free material such as silicon \cite{1009, 1009a},  carbon nanotubes \cite{1011, 1012} or graphene quantum dots\cite{1010}, or to store the quantum bit in a spin state of the nitrogen vacancy center in diamond\cite{1110a,1110,1111,1112}.

Recently the spin state of the hole emerged as an alternative and very promising candidate for the realization of a qubit\cite{qd28,qd291,qd292} in semiconductor solid state systems. Its main advantage over the electron spin is the fact that the hole is less sensitive to the interaction with the nuclear spin of the surrounding material. Since the hole is described by a p-type orbital in many semiconductors, its wave function vanishes at the nuclear site and thus the contact hyperfine interaction between hole spin and nuclear spin is canceled. Even though holes still experience interaction with nuclear spins with dipolar character, it is about ten times weaker than the contact interaction for electrons\cite{qd25,qd52,qd53,qd53a,qd55d,qd17}. Consequently, the coherence time of the spin state of the hole is longer than for the electron spin. The coherence time also depends on the heavy hole (HH)-light hole (LH) mixing. For pure HH states, the coherence time of the hole reaches its maximum because the interaction between hole spin and nuclear spins has an Ising type character\cite{qd25,qd53}.

Despite the fact that many experimental and theoretical investigations have been done on hole spin relaxation and decoherence mechanisms \cite{qd25,qd52,qd53,qd53a,qd55d,qd17,HH-LH_m0,HH-LH_m1,qd23,qd26,qd88,qd55,qd30,1002,1007,1008,1021,qd100,1018,1019}, so far hole spin dynamics in semiconductor nanostructures is still largely unexplored and needs deeper understanding. However, the fact that holes are alternative long living qubits has stimulated progress in the experimental realization of hole spin preparation, manipulation, and read out\cite{qd18,qd54,qd55a,1003,qd55b,qd55c,qd55d,qd55e,qd55f}. It is quite remarkable that it is even possible to initialize hole spin states with very high fidelity ($99\%$) without the application of an external magnetic field \cite{qd26}. Very recently electrical control of a single hole spin in a gated InSb nanowire has been realized \cite{qd56}. Other theoretical proposals for hole spin control are EDSR (electron dipole spin resonance) techniques for heavy holes \cite{qd32}, non-Abelian geometric phases \cite{qd32aa}, the application of a static magnetic field applied in quantum dots \cite{qd31}, and an electric g tensor manipulation \cite{qd32bb, qd32cc} and are waiting for their experimental realization.

  Another important and indispensable aspect for the realization of a quantum computer architecture is scalability.
Recently, scalable architectures were proposed where long distant qubit coupling might be obtained via floating gates\cite{prx}. Furthermore, coupling between spin qubits defined in a semiconductor InAs nanowire and a superconducting cavity\cite{1005} was experimentally realized which is particularly promising for future realizations of scalable networks of spin qubits. A scalable architecture for optically controlled hole spin qubits confined in quantum dot molecules was also proposed \cite{1001}.

   Recently we have shown that the motion of a hole in gated semiconductor nanodevices can induce heavy hole spin rotations in the presence of the  Dresselhaus spin orbit coupling (DSOI) \cite{qdprl}. We proposed a nanodevice based on GaAs which can act as a  quantum NOT (Pauli $X$) gate. In this paper we propose a couple of nanodevices capable to realize other single quantum logic gates: Pauli $Y$ and $Z$ gates and a $U_S$ gate which can realize a balanced superposition of qubit basis states. The required quantum logic operation is realized by transporting the hole around a rectangular loop which is defined by metal electrodes which cover the semiconductor nanostructure. The geometry of the metal gates determines the hole trajectory and consequently the type of quantum operation which we want to perform. Moreover, we propose a so called combo nanodevice in which each of the proposed quantum logic gates (Pauli $X$, $Y$, $Z$ and $U_S$) can be applied in an arbitrary sequence on a HH spin qubit. We give a full theoretical description of the nanodevices and present the results of time dependent simulations. The description of the all electrical control scheme which has to be applied in order to perform the desired quantum gate by the proposed nanodevice is provided.  Moreover, thanks to the fact that the proposed gates are only controlled by weak static voltages applied to the local top electrodes, it is possible to realize a scalable quantum architecture in which each qubit can be addressed individually without disturbing the state of other qubits in the quantum register.

In this paper we perform our simulations for CdTe, and not GaAs as in Ref.~\onlinecite{qdprl} for several reasons. Due to the smaller dielectric constant and higher in plane effective mass, the binding energy of a self trapped hole under a metal gate in a CdTe quantum well is larger and consequently the hole soliton effect is more pronounced than in the previously used GaAs material. Since the Cd and Te isotopes are characterized by a nuclear spin $I=\frac{1}{2}$, the dipolar hyperfine interaction between the hole spin and the nuclear spin of the host material is weaker than for GaAs which nuclei have spin $I=\frac{3}{2}$. Furthermore, the dephasing time of an electron or a hole confined in a quantum dot made from II-VI group compounds is a few times longer than for III-V compounds because of the significantly lower natural concentration of isotopes with nonzero nuclear magnetic moment (Ga $100\%$, As $100\%$, Cd $25\%$, Te $7.8\%$) \cite{qd60,qd53,II-VI}.

The lateral size of proposed nanodevices is determined by the $\lambda_{SO}$ length: the distance which has to be traveled by the hole in order to perform a full $2\pi$ HH spin rotation. Since $\lambda_{SO}^{GaAs}\approx 4000$nm and $\lambda_{SO}^{CdTe}\approx 700$nm the proposed nanodevices which are based on CdTe are significantly smaller than those based on GaAs.

The proposed devices can also be realized in other zinc-blende semiconductors, but due to different material parameters they will differ in size and gate operation time\cite{qdprl}.

   This paper is further organized as follows. Section II describes the general device layout and discusses the applied theoretical model, i.e. our self consistent Poisson-Schr\"{o}dinger approach with the Luttinger-Kohn Hamiltonian. The ground state wave functions are presented in Section III. In Section IV we present and describe the separate nanodevices acting as quantum logic gates on heavy hole spin states, together with the results of our time-dependent simulations. The combo nanodevice in which an arbitrary sequence of single quantum logic gates can be performed on a HH spin state is described in Section VI as well as the proposal of a scalable architecture. Section VI summarizes the obtained results.

\indent
\section{DEVICE AND THEORETICAL MODEL}
\begin{figure}[ht!]
\epsfxsize=70mm \epsfbox[15 306 579 535]{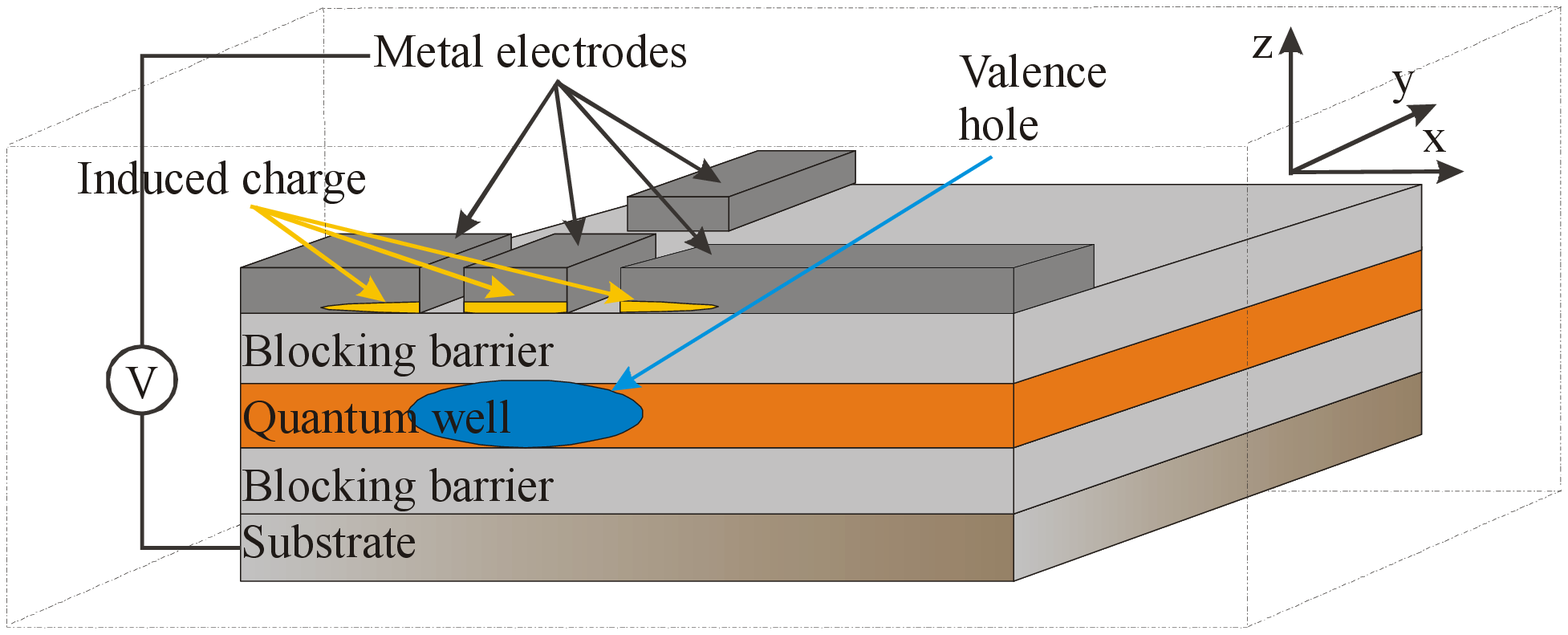}
\caption{Crossection of the nanodevice.\label{NAN}}
\end{figure}

Let us consider a planar semiconductor heterostructure covered by nanostructured metal gates. The system contains a zinc-blende semiconductor quantum well (QW) which is sandwiched between two $10$nm blocking barriers (Fig. \ref{NAN}). The single valence hole is confined in the quantum well region which is oriented in the $z[001]$ (growth) direction and thus the hole can only move in the $x[100]-y[010]$ plane.
   In such a structure, the hole induces a response potential in the electron gas in the metallic gate which in turn leads to a lateral self-confinement of the hole wave function\cite{qd11,qd11a}. This self trapped hole has soliton-like properties: it can be transported as a stable wave packet which maintains its shape during motion. Furthermore, it can reflect or pass through obstacles (potential barriers or wells) with $100\%$ probability while conserving its shape. This property can be used to realize on demand transfer of a hole between different locations within the nanodevice (in the area of the quantum well which is under the metal electrodes) by applying static weak voltages to the electrodes only\cite{qd12}.

In  order to describe the presented system we rely on the two dimensional four band HH-LH Hamiltonian:
\begin{equation}
	\label{eq:HAM}
    \hat{H}=\hat{H}_{LK}^{2D}+|e|\phi(x,y,z_0)\hat{I}+\hat{H}_{BIA}^{2D}.
\end{equation}
The HH (LH) states are characterized by the $J_z=\pm 3/2$ ($J_z=\pm 1/2$) projections of total angular momentum on the $z$ axis. The first term is the Luttinger-Kohn Hamiltonian\cite{qdLutt} describing the kinetic energy of the two-dimensional hole, which for unstrained zinc-blende structures can be written in the effective mass approximation as
\begin{equation}
\label{eq:LK}
\hat{H}_{LK}^{2D}= \left( \begin{array}{cccc}
\hat{P}_{h} & 0 & \hat{R} & 0 \\
0 & \hat{P}_{l} & 0 & \hat{R} \\
\hat{R}^{\dag} & 0 & \hat{P}_{l} & 0  \\
0 & \hat{R}^{\dag} & 0 & \hat{P}_{h}  \end{array} \right),
\end{equation}
where
\begin{eqnarray}
\label{eq:ELEMENTS}
\hat{P}_{h}&=&\frac{\hbar^2}{2m_0}(\gamma_1+\gamma_2)(k_x^2+k_y^2)+E_0^+\nonumber\\
\hat{P}_{l}&=&\frac{\hbar^2}{2m_0}(\gamma_1-\gamma_2)(k_x^2+k_y^2)+E_0^-\nonumber\\
\hat{R}&=&\frac{\hbar^2}{2m_0}\sqrt{3}[\gamma_2(k_x^2-k_y^2)-2i\gamma_3k_xk_y]
\end{eqnarray}
We denote $E_0^\pm=\frac{\hbar^2}{2m_0}(\gamma_1\mp 2\gamma_2)\langle k_z^2\rangle $ as the first subband energy in the $z$ direction $(E_0^-=E_\perp^{LH}, E_0^+=E_\perp^{HH})$ with $\langle k_z^2\rangle=\pi^2/d^2$, where $d$ is the quantum well width, $\gamma_1, \gamma_2, \gamma_3$ are the Luttinger parameters and $m_0$ is the free electron mass. The momentum operators are defined as $\hbar k_q=-i \hbar \frac{\partial}{\partial q}$ where $q=x,y$. $\hat{I}$ is the unit operator, $e$ is the elementary charge and $z_0$ is the center of the quantum well. We use the representation where the projections of the Bloch angular momentum on the $z$ axis are arranged in the following order: $J_z=\frac{3}{2}, \frac{1}{2}, -\frac{1}{2}, -\frac{3}{2}$ ($|HH\uparrow\rangle, |LH\uparrow\rangle, |LH\downarrow\rangle, |HH\downarrow\rangle$). Consistently with this convention the state vector can be written as
\begin{equation}
\label{eq:LSPINOR}
\Psi(x,y,t)=\left( \begin{array}{c}
\psi_{HH}^\uparrow(x,y,t)\\
\psi_{LH}^\uparrow(x,y,t)\\
\psi_{LH}^\downarrow(x,y,t)\\
\psi_{HH}^\downarrow(x,y,t) \end{array} \right)
\end{equation}
The electrostatic potential $\phi(x,y,z_0,t)$ which is ``felt" by the hole is the source of the self trapping potential. Its origin is due to charges induced on the metal electrodes. The electrostatic potential $\phi(x,y,z_0,t)$ can be calculated according to the superposition principle and it is the difference between the total electrostatic potential and the self-interaction potential $\phi(x,y,z_0,t)=\Phi_{TOT}(x,y,z_0,t)-\phi_{si}(x,y,z_0,t)$. The total electrostatic potential distribution within the considered system is found by solving the Poisson equation in a three dimensional computational box containing the entire nanodevice:
\begin{equation}
	\label{eq:POISS}
    \nabla^2\Phi_{tot}(x, y, z,t)=-\frac{1}{\epsilon\epsilon_0}\rho_{tot}(x, y,z,t)
\end{equation}
The charge density of a single hole is described by the two-dimensional distribution $\rho_{tot}(x,y,z,t)=\rho(x,y,t)\delta(z-z_0)$, where
\begin{eqnarray}
\label{eq:DENSITY}
   \rho(x,y,t)&=&|e|(|\psi_{HH}^\uparrow(x,y,t)|^2+|\psi_{LH}^\uparrow(x,y,t)|^2 \nonumber \\
&+&|\psi_{LH}^\downarrow(x,y,t)|^2+
|\psi_{HH}^\downarrow(x,y,t)|^2).
\end{eqnarray}

The self interaction potential $\phi_{si}(x,y,z_0,t)$ is directly connected to the total wave packet charge density distribution and can be calculated straightforwardly as follows:
\begin{equation}
\label{eq:SI}
    \phi_{si}({\bf r,t})=\frac{1}{4\pi\epsilon\epsilon_0}\int d{\bf r} \frac{\rho({\bf r}',t)}{|{\bf r}-{\bf r}'|}.
\end{equation}
Quantum calculations \cite{qd13} indicate that electrostatic approach described above is a good approximation of the actual response potential of the electron gas.

Since in the considered system the hole is confined in the zinc-blende semiconductor (thus lacking crystal inversion symmetry) quantum well we have to take into account the DSOI \cite{qd14} described by the $\hat{H}_{BIA}$ Hamiltonian which for holes in bulk (including the two leading contributions) can be written as \cite{qd39}:
\begin{eqnarray}
\label{eq:HBIA}
\nonumber \hat{H}_{BIA}=&-&\beta_0\Big[k_x\{J_x,J_y^2-J_z^2\}+k_y\{J_y,J_z^2-J_x^2\}\\
\nonumber &+&k_z\{J_z,J_x^2-J_y^2\}\Big]\\
\nonumber &-&\beta \Big[\{k_x,k_y^2-k_z^2\}J_x+\{k_y,k_z^2-k_x^2\}J_y\\
&+&\{k_z,k_x^2-k_y^2\}J_z\Big]
\end{eqnarray}
where $\boldsymbol{k}=(k_x, k_y, k_z)$ is the momentum vector and $\boldsymbol{J}=(J_x, J_y, J_z)$ is the vector of the $4\times 4$ spin $3/2$ matrices. We denote half of the anticommutator as $\{{A},{B}\}=\frac{1}{2}({A}{B}+{B}{A})$. Going from bulk to two-dimensional systems and neglecting qubic $k$ terms \cite{Dq} the bulk DSOI can be directly transformed and expressed in the matrix form as
\begin{eqnarray}
\hat{H}_{BIA}^{2D}&=&
\nonumber
-\beta_0\left[ k_x\{J_x,J_y^2-J_z^2\}+k_y\{J_y,J_z^2-J_x^2\}\right]\\
\nonumber&+&\beta \langle k_z^2\rangle(k_xJ_x-k_yJ_y)\\
\nonumber
&=&\frac{\beta_0}{4}\left( \begin{array}{cccc}
0 & \sqrt{3}k_+ & 0 & 3k_- \\
\sqrt{3}k_- & 0 & -3k_+ & 0 \\
0 & -3k_- & 0 & \sqrt{3}k_+  \\
3k_+ & 0 & \sqrt{3}k_- & 0  \end{array} \right)\\
&+&
\nonumber
\frac{\beta \langle k_z^2\rangle}{2}\left( \begin{array}{cccc}
0 & \sqrt{3}k_+ & 0 & 0 \\
\sqrt{3}k_- & 0 & 2k_+ & 0 \\
0 & 2k_- & 0 & \sqrt{3}k_+  \\
0 & 0 & \sqrt{3}k_- & 0  \end{array} \right)\\
\end{eqnarray}
where $k_\pm=k_x\pm ik_y$. Similar as in the Luttinger-Kohn Hamiltonian one has $\hbar k_q=-i\hbar \frac{\partial}{\partial q}$ where $q=x,y$. Numerical estimates of the DSOI coupling constants $\beta_0$ and $\beta$ for different materials can be found in Refs.~\onlinecite{qd39, qd39a}. We assume that the CdTe quantum well is symmetric in the $z$ direction thus Rashba spin orbit interaction is absent in the investigated systems.

The time evolution of the system is described by the time dependent Schr\"{o}dinger equation which is solved numerically in an iterative manner:
\begin{equation}
\label{eq:STIME}
    \Psi(x,y,t+dt)=\Psi(x,y,t-dt)-\frac{2idt}{\hbar}\hat{H}\Psi(x,y,t),
\end{equation}
which is solved self-consistently with the Poisson equation \eqref{eq:POISS} and the self-interaction potential \eqref{eq:SI}. Since the hole wave packet is moving the Poisson equation has to be solved in every time step of the iteration procedure. We take the ground state wave function $\Psi_0(x,y)=\Psi(x,y,t_0)$ of the self confined hole under the metal electrode as initial condition for the time evolution numerical scheme \eqref{eq:STIME}. This ground state wave function is found by solving the stationary Schr\"{o}dinger equation
\begin{equation}
\label{eq:SSTAT}
    \hat{H}\Psi_0(x,y)=E\Psi_0(x,y)
\end{equation}
using the imaginary time propagation (ITP) method \cite{itp}.
\indent
\section
{Ground state wave function}

   It is important to know the contribution of different basis states in the ground state of the self confined hole under the metal electrodes, i.e. the mixing of HH and LH states. We consider the system from Fig. \ref{NAN}: the hole is confined in the CdTe quantum well and covered by the system of electrodes. The center of the QW is $15$nm distant from the top metal electrodes (the QW layer and blocking layers are $10$nm thick). We perform calculations for CdTe with following Luttinger parameters: $\gamma_1^{CdTe}=5.3, \gamma_2^{CdTe}=1.7, \gamma_3^{CdTe}=2.$, dielectric constant $\epsilon^{CdTe}=10.125$ and the DSOI coupling constants $\beta_0=0.027$ eV\AA, $\beta=76.93$ eV\AA$^3$. The hole is initially ``prepared" in the $|HH\uparrow\rangle$ state. After the ITP procedure, the system relaxes to the ``real" hole ground state. Let us now consider two situations, when the DSOI is present and when it is absent in the system.

\begin{figure}[ht!]
\epsfxsize=83mm \epsfbox[16 211 581 632]{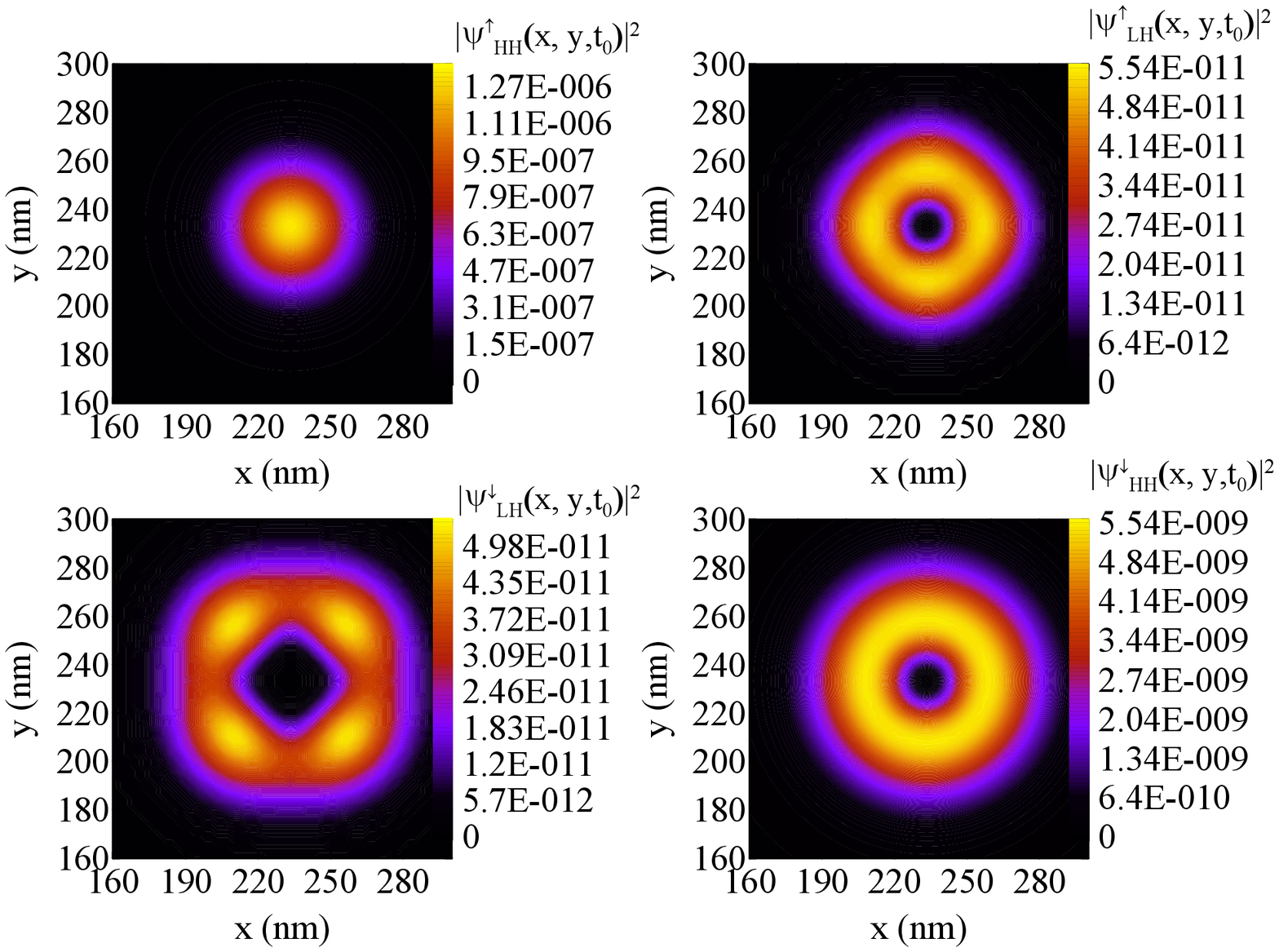}
\caption{ Modulus square of the components of the hole Luttinger spinor wave function: $|\psi_{HH}^{\uparrow}(x,y, t_0)|^2$, $|\psi_{LH}^{\uparrow}(x,y, t_0)|^2$, $|\psi_{LH}^{\downarrow}(x,y, t_0)|^2$ and $|\psi_{HH}^{\downarrow}(x,y, t_0)|^2$ in the spin up ground state in the presence of the DSOI interaction. It can be noticed that $\psi_{HH}^{\uparrow}(x,y, t_0)$ has the biggest contribution to the total wave function $\Psi(x,y, t_0)$, while the LH components $\psi_{LH}^{\uparrow}(x,y, t_0)$, $\psi_{LH}^{\downarrow}(x,y, t_0)$ are about 4 orders of magnitude smaller. In case of absence of the DSOI there are only two non zero components $\psi_{HH}^{\uparrow}(x,y, t_0)$, $\psi_{LH}^{\downarrow}(x,y, t_0)$ and their modulus square looks identical as for those in the presence of DSOI as depicted in the above figure. \label{WF}}
\end{figure}

For non zero DSOI coupling constants we obtain the following probabilities to occupy the different basis hole states:
\begin{eqnarray*}
    P_{|HH\uparrow\rangle}&=&\int dxdy|\psi_{HH}^{\uparrow}(x,y, t_0)|^2\approx 0.99\\
    P_{|HH\downarrow\rangle}&=&\int dxdy|\psi_{HH}^{\downarrow}(x,y, t_0)|^2\approx 0.01\\
P_{|LH\uparrow\rangle}&=&\int dxdy|\psi_{LH}^{\uparrow}(x,y, t_0)|^2\approx 8.6\cdot 10^{-5}\\
P_{|LH\downarrow\rangle}&=&\int dxdy|\psi_{LH}^{\downarrow}(x,y, t_0)|^2\approx 1.1\cdot 10^{-4}
\end{eqnarray*}
The modulus square of the components of the Luttinger spinor ground state wave function $\Psi(x,y, t_0)$ are plotted in Fig. \ref{WF}.
When the DSOI is absent only the $|HH\uparrow\rangle$ and $|HH\downarrow\rangle$ are occupied with following probabilities: $P_{|HH\uparrow\rangle}\approx 1$, $P_{|LH\downarrow\rangle}\approx 8.7\cdot 10^{-5}$. The obtained results show that the mixing between HH and LH states is negligible, the DSOI induces only a very small mixing into the HH state. This is an important result because in systems in which the hole occupies only the HH band, the hole spin coherence time is significantly longer than for electron spin\cite{qd25,qd53}. Furthermore, we can say that in the considered system the hole spin qubit is well defined with $99\%$ probability in the subspace of the HH spin basis states.

\section{QUANTUM GATES}
   Recently we have shown that the motion of the hole along an induced quantum wires in the presence of DSOI can induce HH spin rotations \cite{qdprl}. In particular, during the motion of the hole along $x$ ($[100]$) and $y$ ($[010]$) direction, its spin rotates (precesses) around the axis parallel to the direction of motion and this process can be associated with following operations: $\hat{R}_x(\phi)$ and $\hat{R}_y(\phi)$. Their explicit form is\cite{op}:
\begin{equation}
\label{eq:RX}
    \hat{R}_x(\phi)=\frac{1}{\sqrt{2}}\left( \begin{array}{cc}
i\sqrt{1+\cos(\phi)} & \frac{\sin(\phi)}{\sqrt{1+\cos(\phi)}} \\
\frac{\sin(\phi)}{\sqrt{1+\cos(\phi)}} & i\sqrt{1+\cos(\phi)}\end{array} \right)
\end{equation}

\begin{equation}
\label{eq:RY}
    \hat{R}_y(\phi)=\frac{1}{\sqrt{2}}\left( \begin{array}{cc}
\sqrt{1+\cos(\phi)} & -\frac{\sin(\phi)}{\sqrt{1+\cos(\phi)}} \\
\frac{\sin(\phi)}{\sqrt{1+\cos(\phi)}} & \sqrt{1+\cos(\phi)}\end{array} \right),
\end{equation}
where $\phi(t) =2\pi\frac{\lambda(t)}{\lambda_{SO}}$ is the rotation angle while $\lambda(t)$ is the distance traveled by the hole after time $t$, $q=x$, $y$ is the direction of motion as well as the axis around which the HH spin is rotated. After passing the distance $\lambda_{SO}$, the HH spin makes a full $2\pi$ rotation. The above operators \eqref{eq:RX},\eqref{eq:RY} act on the following wave function:
\begin{equation}
    \Psi_{HH}(x,y,t)=\left( \begin{array}{c}
\psi_{HH}^\uparrow(x,y,t)\\
\psi_{HH}^\downarrow(x,y,t) \end{array} \right)
\end{equation}
which is defined in the subspace of HH basis states. For such a wave function we define the expectation value of the HH pseudo-spin $\frac{1}{2}$ as $s_i(t)=\frac{3}{2}\hbar\langle\Psi_{HH}(x,y,t)|\sigma_i|\Psi_{HH}(x,y,t)\rangle$, where the $\sigma_i$ is a Pauli matrix, $i=x, y, z$.

  Taking advantage of the fact that hole motion induces HH spin rotations, we can design nanodevices which are able to realize various single quantum logic gates. We use operators \eqref{eq:RX},\eqref{eq:RY} to determine the topology of the metal electrodes that cover the nanodevice and in this way determine the hole trajectory which is passed by the hole during the realization of a certain quantum gate on a HH spin qubit. We propose nanodevices which can act as a quantum Pauli $X, Y,$ and $Z$ gate:
\begin{equation}
    \hat{\sigma}_x=\left( \begin{array}{cc}
0 & 1 \\
1 & 0 \end{array}\right),
\hat{\sigma}_y=\left( \begin{array}{cc}
0 & -i \\
i & 0 \end{array}\right),
\hat{\sigma}_z=\left( \begin{array}{cc}
1 & 0 \\
0 & -1 \end{array}\right).
\end{equation}
Furthermore we propose a nanodevice which is able to perform a quantum logic operation similar to the Hadamard gate, which we call the $U_S$ gate:
\begin{equation}
\hat{U}_{S}=\frac{1}{\sqrt{2}}\left( \begin{array}{cc}
-1 & i \\
1 & i \end{array}\right), \: \: \:
\hat{U}_{S}^{-1}=\frac{1}{\sqrt{2}}\left( \begin{array}{cc}
1 & -1 \\
i & i \end{array}\right).
\end{equation}
The Pauli $Q$ gate performs the HH spin rotation about an angle $\pi$ around the Q axis where $Q=X, Y, Z$. The $s_k=\frac{3}{2}\hbar$ HH spin state can be transformed into the $s_k=-\frac{3}{2}\hbar$ state using the $\sigma_i$ or $\sigma_j$ gate where $i, j, k$ can take $x, y, z$ values while $i\neq j\neq k$.
\begin{figure}[ht!]
\epsfxsize=70mm \epsfbox[72 41 527 820]{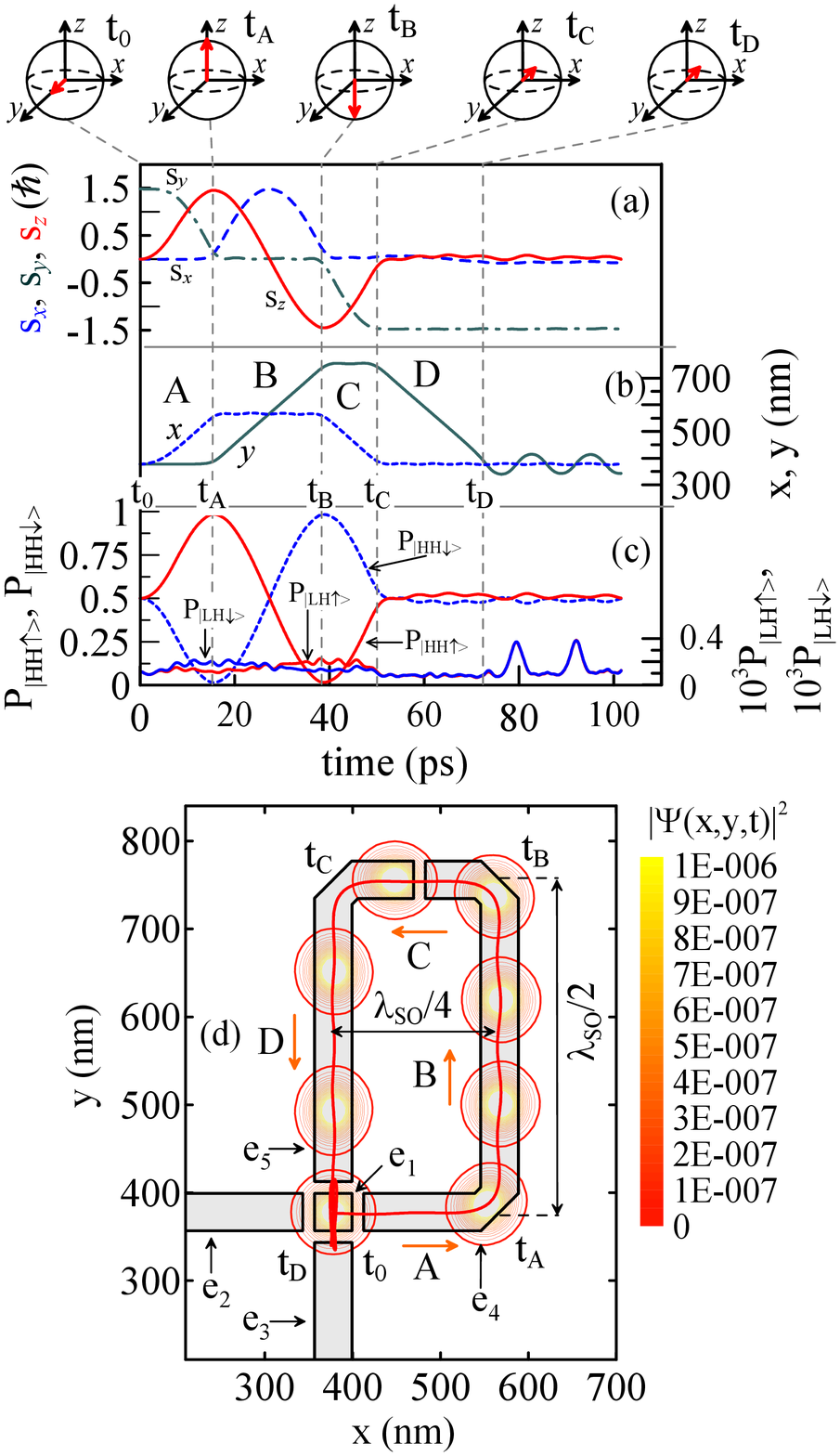}
\caption{The time evolution of the expectation value of the HH spin components $s_x(t), s_y(t), s_z(t)$ (a), average position $x(t), y(t)$ of the hole wave packet (b) and the occupation probabilities $P_{|HH\uparrow\rangle}(t), P_{|HH\downarrow\rangle}(t), P_{|LH\uparrow\rangle}(t), P_{|LH\downarrow\rangle}(t)$ of the hole basis states (c), for the quantum Pauli $X$ (NOT) gate which is covered by the system of electrodes $e_1-e_5$ presented in (d).
In figure (c) the left (right) axis corresponds to the probability of finding the hole in the HH (LH) spin states. In figure (d) the solid red line represents the hole trajectory (the orange arrow represents the direction of motion of the hole). The hole is initially confined under electrode $e_1$ and moves in the $+x$ direction.
 The HH spin qubit state is depicted on the Bloch spheres at times $t_0$, $t_A$, $t_B$, $t_C$, $t_D$ of the quantum gate operation cycle. The contour plots represent the charge density $\rho(x,y,t)$ at a few selected moments in time. \label{PX}}

\end{figure}

The easiest to design and to implement (within the proposed nanostructure) quantum gates that transform one basis state into a balanced superposition of two basis states of the qubit are the $U_S$ and $U_S^{-1}$ gates. Their functionality is similar to the Hadamard gate but since $U_S^2\neq I_{2\times 2}$ ($U_S^3=I_{2\times 2}$) $U_S$ is not exactly a Hadamard gate. Application of the $U_S$ gate is equivalent to the rotation of the $s_{x}=\pm\frac{3}{2}\hbar$, $s_{z}=\pm\frac{3}{2}\hbar$ ($s_{y}=\pm\frac{3}{2}\hbar$) HH spin states around the $z, y,$ ($x$) axis about an angle $\pi/2$, $\pi/2$ ($-\pi/2$), respectively, such that the states $s_{y}=\mp\frac{3}{2}\hbar$, $s_{x}=\mp\frac{3}{2}\hbar$ ($s_{z}=\pm\frac{3}{2}\hbar$) are produced. The reverse process can be obtained by applying the $U_S^{-1}$ gate.

\begin{figure}[ht!]

\epsfxsize=70mm \epsfbox[56 41 545 822]{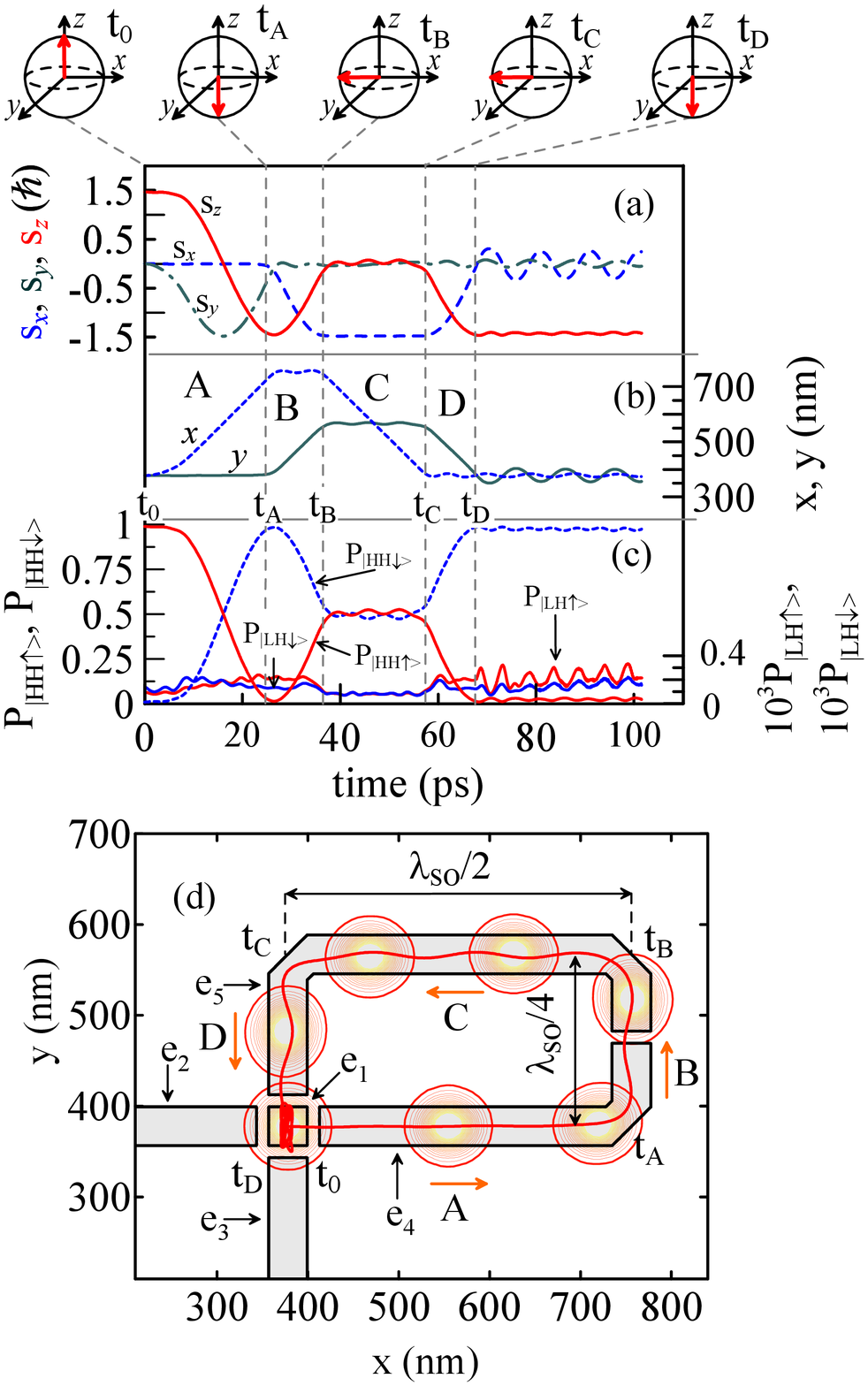}
\caption{The same as Fig. \ref{PX}, but for the quantum Pauli $Y$ gate which is covered by the system of electrodes $e_1-e_5$ presented in (d). \label{PY}}
\end{figure}

\begin{figure}[ht!]
\epsfxsize=70mm \epsfbox[63 48 552 820]{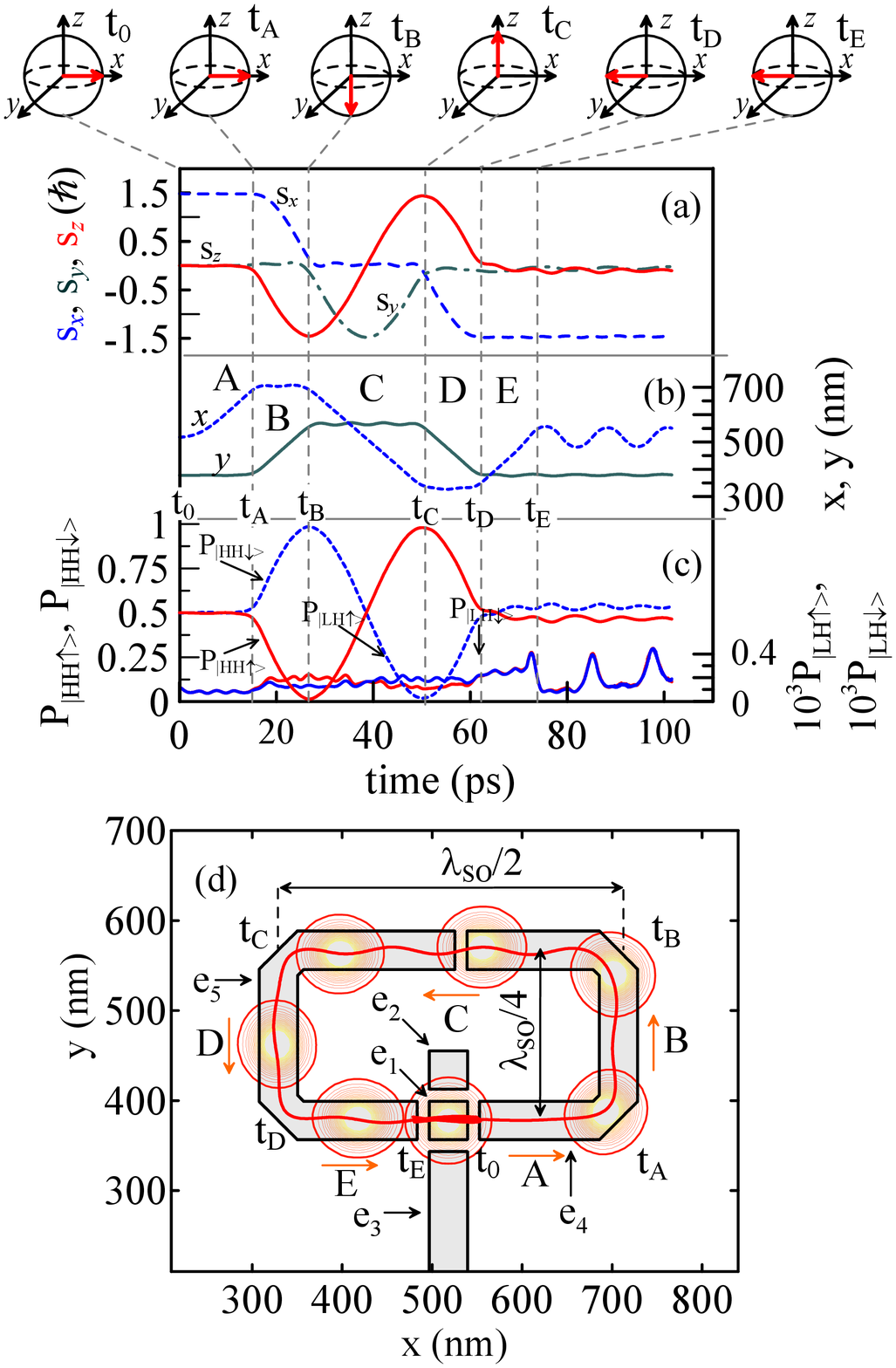}
\caption{ The same as Fig. \ref{PX}, but for the quantum Pauli $Z$ gate which is covered by the system of electrodes $e_1-e_5$ presented in (d). \label{PZ}}
\end{figure}

\begin{figure}[ht!]
\epsfxsize=70mm \epsfbox[62 10 536 820]{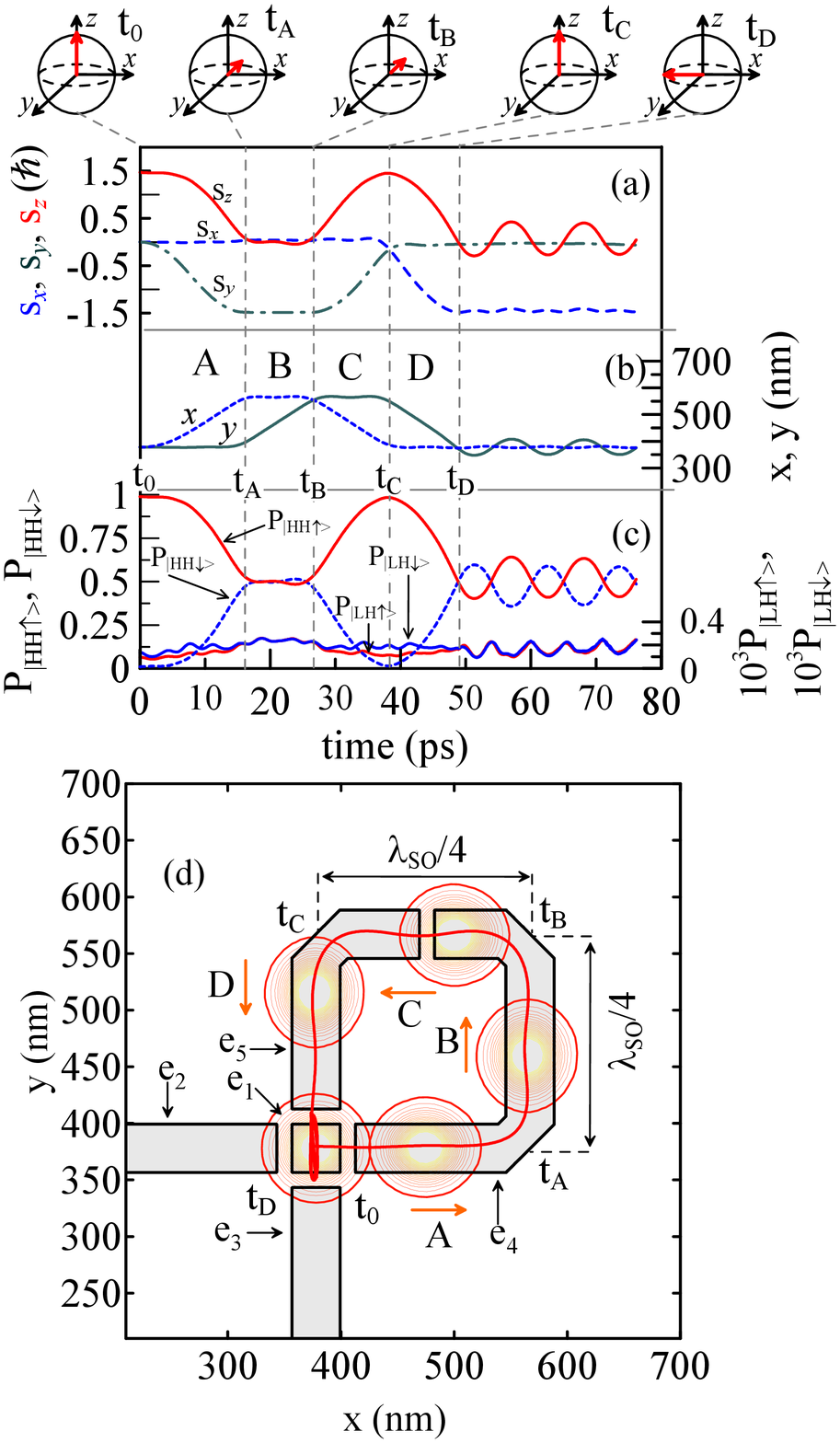}
\caption{ The same as Fig. \ref{PX}, but for the quantum $U^{S}$ gate which is covered by the system of electrodes $e_1-e_5$ presented in (d). \label{US}}
\end{figure}

In order to demonstrate how quantum logic operations are realized we make a precise numerical time dependent simulation. We depict the time evolution of the expectation value of the HH spin $s_x(t)$, $s_y(t)$, $s_z(t)$, the average position of the hole $x(t)$, $y(t)$ and the probability of occupying the hole basis states $|HH\uparrow \rangle$, $|LH\uparrow \rangle$, $|LH\downarrow \rangle$, $|HH\downarrow \rangle$ in parts (a), (b), and (c) of Figs. \ref{PX},\ref{PY},\ref{PZ},\ref{US} for each quantum operation process. The nanodevices are covered by a specially designed system of electrodes which define the path - a closed rectangular loop - which has to be traveled by the hole in order to realize the desired quantum logic operation. The scheme of metal electrodes labeled by $e_1-e_5$ which cover the nanodevices, the hole trajectory and the contour plots of the hole charge density at a few moments of time are depicted in part (d) of Figs. \ref{PX},\ref{PY},\ref{PZ},\ref{US}.

In the initial step of each quantum operation process, the hole is confined under electrode $e_1$ with dimensions $50 \times 50$nm on which a constant $V_1=-0.3$mV voltage is applied. The voltage applied to the other electrodes $e_{2,3,4,5}$ is set to $V_{2, 3, 4, 5}=0$. The distance between $e_1$ and the neighbor $e_{2,3,4,5}$ electrodes is about $7$nm. It should be mentioned that due to the Schotky contact, the Schotky voltage $V_{Schotky}$ has to be taken into account with mV accuracy and the ``real" voltage applied to the metal gates is $V_i\rightarrow V_i-V_{Schotky}$. $V_{Schotky}$ should be determined experimentally for a particular structure. In case of the  Pauli $X$ (NOT) gate we assume that the hole is initially prepared in the HH spin $s_y=\frac{3}{2}\hbar$ state:
\begin{equation}
\Psi(x,y,t_0)=\frac{1}{\sqrt{2}}\left( \begin{array}{c}
\psi_{HH}^\uparrow(x,y,t_0)\\
0\\
0\\
i\psi_{HH}^\downarrow(x,y,t_0) \end{array} \right),
\end{equation}
while the ``magnetic free" preparation of the hole spin states can be achieved using experimentally demonstrated methods \cite{qd26} or by utilizing an analogous device to those which we recently proposed \cite{qd40} to prepare the electron spin in a certain state without application of a magnetic field.

The hole is forced to move in the $+x$ direction by changing the voltage applied on $e_1$ to $V_1=0$ and switching the voltage on $e_4$ to $V_4=-0.3$mV. In our numerical scheme the voltage is changed linearly in time in a duration of $t_{rise}=0.1$ ps.(For a longer $t_{rise}=1$ps the gate operation time is identical while for $t_{rise}=5$ps the gate operation time is about $2.5$ps longer. It is caused by the slightly smaller initial hole momentum.) After traveling the $\lambda_{SO}/4$ long segment A of the loop the $\hat{R}_x(\pi/2)$ operation is performed on the HH spin. At the end of segment A, the hole wave packet reflects from the potential barrier at the corner of electrode $e_4$ and changes its direction of motion into the $+y$ direction. Next the hole passes the segment B whose length is $\lambda_{SO}/2$ and realizes a $\hat{R}_y(\pi)$ rotation. In the meantime the voltage applied to electrode $e_5$ was set to the voltage of the $e_4$ electrode so that the hole can enter easily under $e_5$. Then the hole passes segments C and D, realizing $\hat{R}_x(-\pi/2)$ and $\hat{R}_y(-\pi)$ operations, respectively, and finally returns under electrode $e_1$ whose voltage is set to $V_1=-0.3$mV, while the voltage on the neighbor electrodes $e_2-e_5$ is set to $V_{2,3,4,5}=0.6$mV. After passing the whole loop, a set of HH spin rotations is performed resulting in the Pauli $X$ operation:
\begin{equation}
\hat{R}_y(-\pi)\hat{R}_x(-\pi/2)\hat{R}_y(\pi)\hat{R}_x(\pi/2)=e^{i3\pi/2}\hat{\sigma}_x
\end{equation}

The Pauli $Y$ gate is realized by the nanodevice covered by the system of electrodes shown in Fig. \ref{PY}(d). In this case, as initial condition in our simulation, we take a HH spin up state $s_z=\frac{3}{2}\hbar$:
\begin{equation}
\Psi(x,y,t_0)=\left( \begin{array}{c}
\psi_{HH}^\uparrow(x,y,t_0)\\
0\\
0\\
0 \end{array} \right).
\end{equation}
At the beginning of the gate operation process the hole is forced to move in the $+x$ direction and follows the trajectory defined by the metal gates deposited on top of the nanodevice. The hole passes the A, B, C, and D segments, realizing appropriate rotations and finally the Pauli $Y$ gate is performed:
\begin{equation}
\hat{R}_y(-\pi/2)\hat{R}_x(-\pi)\hat{R}_y(\pi/2)\hat{R}_x(\pi)=e^{i\pi/2}\hat{\sigma}_y.
\end{equation}

The scheme of the electrodes which cover the nanodevice that acts as a Pauli $Z$  (a phase $\pi$ flip) gate is depicted in Fig. \ref{PZ}(d). Let us assume that initially the hole is prepared in the HH spin $s_x=3/2\hbar$ state:
\begin{equation}
\Psi(x,y,t_0)=\frac{1}{\sqrt{2}}\left( \begin{array}{c}
\psi_{HH}^\uparrow(x,y,t_0)\\
0\\
0\\
\psi_{HH}^\downarrow(x,y,t_0) \end{array} \right).
\end{equation}
After changing gate the voltage configuration to $V_4=-0.3$mV the hole starts to move in the $+x$ direction and subsequently passes A, B, C, D and E segments of the loop and eventually realizes the quantum logic operation
\begin{equation}
\hat{R}_x(\pi/2)\hat{R}_y(-\pi/2)\hat{R}_x(-\pi)\hat{R}_y(\pi/2)\hat{R}_x(\pi/2)=\hat{\sigma}_z.
\end{equation}
The last proposed gate $U_S$ can be realized by the nanodevice which is covered by the system of metal gates shown in Fig. \ref{US}(d). We make a numerical simulation starting with a HH spin up state.
In the first step of this proposal, the hole is injected under the electrode $e_4$ in the $+x$ direction. Then the hole moves along the loop which consist of the segments A, B, C, and D and carries out certain HH spin rotations. Finally, the hole returns to its initial position and the $U_S$ operation is accomplished:
\begin{equation}
\hat{R}_y(-\pi/2)\hat{R}_x(-\pi/2)\hat{R}_y(\pi/2)\hat{R}_x(\pi/2)=e^{i\pi/4}\hat{U}_{S}
\end{equation}
The inverse operation $U_S^{-1}$
\begin{equation}
\hat{R}_x(-\pi/2)\hat{R}_y(-\pi/2)\hat{R}_x(\pi/2)\hat{R}_y(\pi/2)=e^{i\pi/4}\hat{U}_{S}^{-1}
\end{equation}
can be obtained by transporting the hole in the same loop but in the opposite direction.

In all proposed gates the hole returns to its initial position after completing the set of transformations and consequently the quantum logic operation is performed exclusively on the HH spin state. The hole is trapped when it reaches the area under the $e_1$ electrode which can be achieved by applying the following voltage configuration scheme: $e_1=-0.3$mV and $e_{2,3,4,5}=+0.6$mV. Since there is no energy dissipation term in the Hamiltonian \eqref{eq:HAM}, the kinetic energy of the hole (which was transferred to it at the initial time step of the gate operation process) is still present in the system after its trapping. This is the reason why the position of the hole wave packet and the expectation value of its spin oscillate after trapping under the $e_1$ electrode. In general, due to interactions with phonons kinetic energy can be lost and eventually the hole will stop as well as its spin will end its oscillation. The presence of an additional quantum well may also lead to the energy dissipation of the soliton $\cite{qd13}$ caused by the retardation effect. Thus in the presented setup energy dissipation (which not lead to spin dephasing) is rather a desired effect.

Despite the fact that after trapping the hole position still oscillates, in certain cases it may practically not affect the final value of the spin. This can be achieved if in the last step of the gate operation process the hole spin is parallel to its direction of motion like in the case of Pauli $X$ and Pauli $Z$ gates acting on $s_y=\pm \frac{3}{2}\hbar$ and $s_x=\pm \frac{3}{2}\hbar$, respectively.

It should be noticed that in order to achieve a straight hole trajectory after reflection from a corner in the loop, the initial hole velocity (controlled by the magnitude of the voltages) should be properly adjusted. If the voltage is not properly adjusted, the trajectory is oscillating but fortunately it only slightly affects the final value of the spin. This deviation from the perfect gate result is a measure for the gate fidelity $\mathcal{F}_{gate}=|\langle\Psi|U^{\dag}_{perfect}U_{simulated}|\Psi\rangle|^2$. The fidelity of the proposed gates -$\mathcal{F}_{gate}$ - which is slightly affected by these oscillations takes the following values: $98.6\%<\mathcal{F}_{Paili X}<99.4\%$, $99.3\%<\mathcal{F}_{Paili Y}<99.8\%$, $99.7\%<\mathcal{F}_{Paili Z}<99.9\%$, $97.8\%<\mathcal{F}_{U_s}<99.9\%$. The easiest factor to tune which affects the hole trajectory is the initial gate voltage. By properly adjusting this voltage, one can get a straight hole trajectory. From the other hand, with higher voltages the hole moves faster and one can get faster gates but with a slightly smaller gate fidelity.
\section{GATED COMBO NANODEVICE}

\begin{figure}[ht!]
\epsfxsize=70mm \epsfbox[76 12 520 830]{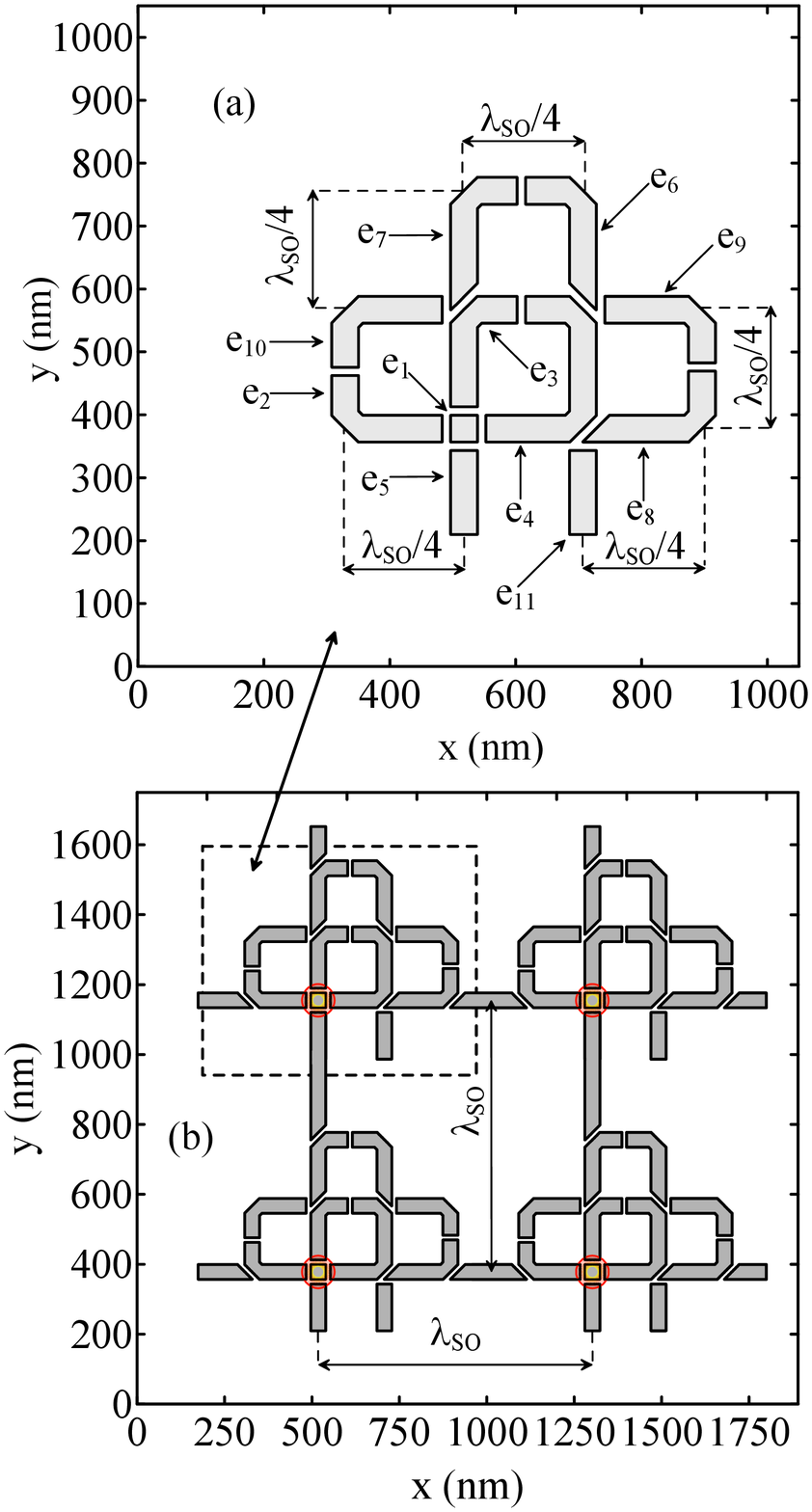}
\caption{The system of electrodes which covers the combo nanodevice (a). The electrodes are labeled with $e_1-e_{11}$. The inter electrode distance is about $7$nm. Fragment of the scalable architecture (b) consisting of $4$ HH spin qubits on which the proposed quantum gates can be applied one by one and in an arbitrary sequence.\label{COMBO_SCAL}
}
\end{figure}

All of the previously proposed nanodevices which realize HH qubit quantum gates can be integrated into a single so called gated combo nanodevice. This device is capable of realizing Pauli $X, Y, Z$ and $U_S$ quantum logic operations in an arbitrary sequence. The nanodevice is covered by $11$ electrodes labeled by $e_1-e_{11}$ which are depicted in Fig. \ref{COMBO_SCAL} (a). 
In order to realize a certain quantum logic gate in this nanodevice a special scheme of voltages $V_1-V_{11}$ has to be applied to the electrodes $e_1-e_{11}$. The voltages have to be switched several times during the gate operation process. We denote $t_0$ as the initial time step at which the hole is confined under electrode $e_1$, which can be achieved by application of the following voltage configuration $V_1=V_0$(in numerical simulation we take $V_0=-0.4$mV) and $V_{2,3,4,5} = 0$ respectively to electrode $e_1$ and its neighbor electrodes $e_{2,3,4,5}$. The time $t_{start}$ corresponds to the moment the hole is forced to move. In all proposed gates (except the $U_{S}^{-1}$ gate) the hole is initially injected from under $e_1$ to under $e_4$ ($e_3$) which is realized by switching the voltage to $V_1=0$ and $V_4=V_0$ ($V_3=V_0$). During the gate operation process the voltage on some electrodes has to be changed at time $t_{change}$ so the hole can enter the appropriate area of the nanodevice. At the end of the gate operation cycle the hole returns to its initial position under $e_1$. At $t_{stop}$ it is captured again by using the following voltage configuration scheme: $V_1=V_0$ and $V_{2,3,4,5} = -2V_0$. The voltage configuration scheme which has to be applied to the electrodes in order to realize a particular quantum logic gate is shown in Table 1.

\begin{figure*}[t!]
\epsfxsize=160mm \epsfbox[16 107 580 739]{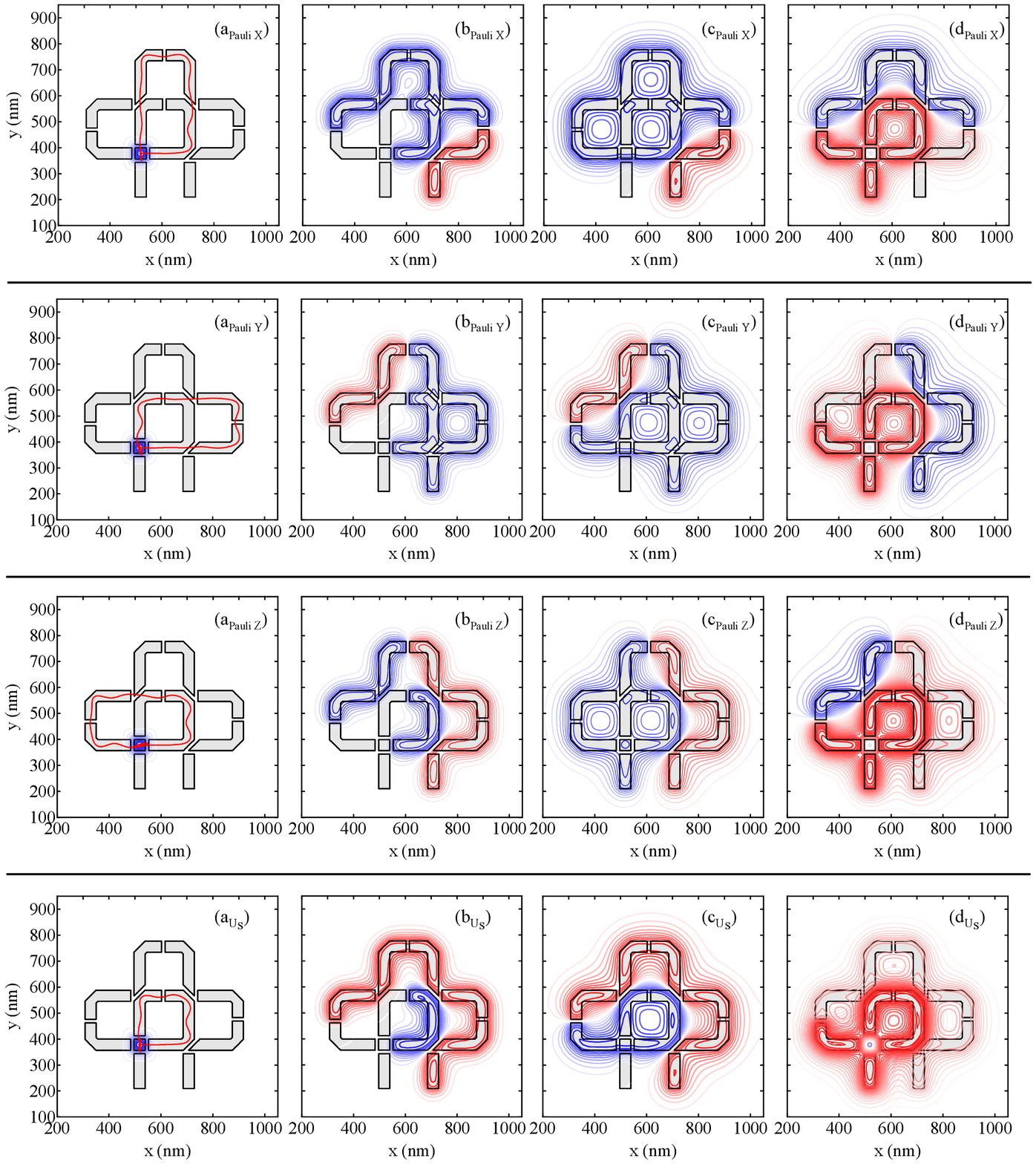}
\caption{The electrostatic potential distribution $\phi_0(x, y, z_0)$ in the quantum well layer $(z_0)$ which comes from the presence of the electrodes to which a certain voltage $V_i$ is applied according to the scheme from Table 1, where $i=1, \ldots ,11$.
The $\phi_0(x, y, z_0)$ is plotted for four crucial moments of time of the quantum gate operation process: ($a_j$) $t\leq t_0$, ($b_j$) $t_{start}<t<t_{change}$, ($c_j$) $t_{change}<t<t_{stop}$, ($d_j$) $t_{stop}<t$. The hole trajectory is depicted in ($a_j$) where $j$ denotes a particular quantum gate. The red isolines denote a positive electrostatic potential distribution while the blue lines a negative one.\label{E_POT}}
\end{figure*}

\begin{table*}[t]
\centering
\begin{tabular}{c||c|c|c|c||c|c|c|c||c|c|c|c||c|c|c|c}
 & \multicolumn{4}{c||}{Pauli $X$ gate} & \multicolumn{4}{|c||}{Pauli $Y$ gate} & \multicolumn{4}{|c||}{Pauli $Z$ gate} & \multicolumn{4}{|c}{$U_S$ gate}\\ \hline
Gate label / time & $t_0$ & $t_{start}$ & $t_{change}$ & $t_{stop}$
                  & $t_0$ & $t_{start}$ & $t_{change}$ & $t_{stop}$
                  & $t_0$ & $t_{start}$ & $t_{change}$ & $t_{stop}$
                  & $t_0$ & $t_{start}$ & $t_{change}$ & $t_{stop}$\\\hline\hline
$V_1$  & $V_0$ & 0 & $V_0$ & $V_0$
             & $V_0$ & 0 & $V_0$ & $V_0$
             & $V_0$ & 0 & $V_0$ & $V_0$
             & $V_0$ & 0 & $V_0$ & $V_0$ \\\hline
$V_2$  & 0 & 0 & $V_0$ & $-2V_0$
             & 0 & 0 & $V_0$ & $-2V_0$
             & 0 & 0 & $V_0$ & $-2V_0$
             & 0 & 0 & $V_0$ & $-2V_0$ \\\hline
$V_3$  & 0 & 0 & $V_0$ & $-2V_0$
             & 0 & 0 & $V_0$ & $-2V_0$
             & 0 & 0 & $V_0$ & $-2V_0$
             & 0 & 0 & $V_0$ & $-2V_0$ \\\hline
$V_4$  & 0 & $V_0$ & $V_0$ & $-2V_0$
             & 0 & $V_0$ & $V_0$ & $-2V_0$
             & 0 & $V_0$ & $V_0$ & $-2V_0$
             & 0 & $V_0$ & $V_0$ & $-2V_0$ \\\hline
$V_5$  & 0 & 0 & 0 & $-2V_0$
             & 0 & 0 & 0 & $-2V_0$
             & 0 & 0 & $V_0$ & $-2V_0$
             & 0 & 0 & 0 & $-2V_0$ \\\hline
$V_6$  & 0 & $V_0$ & $V_0$ & $V_0$
             & 0 & $V_0$ & $V_0$ & $V_0$
             & 0 & $-V_0$ & $-V_0$ & $-V_0$
             & 0 & $-V_0$ & $-V_0$ & $-V_0$ \\\hline
$V_7$  & 0 & $V_0$ & $V_0$ & $V_0$
             & 0 & $-V_0$ & $-V_0$ & $-V_0$
             & 0 & $V_0$ & $V_0$ & $V_0$
             & 0 & $-V_0$ & $-V_0$ & $-V_0$ \\\hline
$V_8$  & 0 & $-V_0$ & $-V_0$ & $-V_0$
             & 0 & $V_0$ & $V_0$ & $V_0$
             & 0 & $-V_0$ & $-V_0$ & $-V_0$
             & 0 & $-V_0$ & $-V_0$ & $-V_0$ \\\hline
$V_9$ & 0 & $V_0$ & $V_0$ & $V_0$
             & 0 & $V_0$ & $V_0$ & $V_0$
             & 0 & $-V_0$ & $-V_0$ & $-V_0$
             & 0 & $-V_0$ & $-V_0$ & $-V_0$ \\\hline
$V_{10}$ & 0 & $V_0$ & $V_0$ & $V_0$
                & 0 & $-V_0$ & $-V_0$ & $-V_0$
                & 0 & $V_0$ & $V_0$ & $V_0$
                & 0 & $-V_0$ & $-V_0$ & $-V_0$ \\\hline
$V_{11}$ & 0 & $-V_0$ & $-V_0$ & $-V_0$
                & 0 & $V_0$ & $V_0$ & $V_0$
                & 0 & $-V_0$ & $-V_0$ & $-V_0$
                & 0 & $-V_0$ & $-V_0$ & $-V_0$ \\ 
\end{tabular}
\indent

\caption{Proposed voltage configuration scheme which has to be applied to the electrodes that cover the gated combo nanodevice in order to realize a Pauli $X, Y, Z$ and $U_S$ quantum logic operation. In the presented simulation we take $V_0=-0.4$mV. }
\label{tab:1}
\end{table*}
\begin{figure*}[ht!]
\epsfxsize=183mm \epsfbox[17 350 567 490]{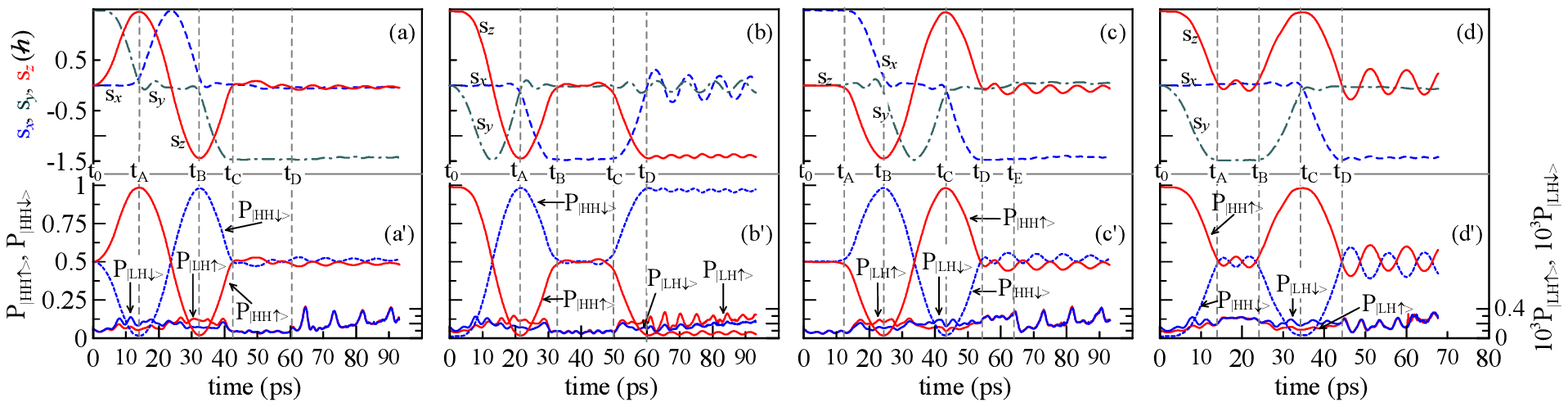}
\caption{Time evolution of the HH spin components for the Pauli $X$ (a), $Y$ (b), $Z$ (c) and $U_S$ gate (d) realized by the gated combo nanodevice which is covered by the system of electrodes shown in Fig. \ref{COMBO_SCAL} (a). The electrode voltage scheme which is responsible for initialization and control of a particular quantum logic operation can be found in Table 1. 
The corresponding occupation probabilities for the HH and LH spin states are depicted in (a'), (b'), (c'), (d'). The hole trajectory for each quantum gate realized by the combo nanodevice can be found in Fig. \ref{E_POT}($a_j$), where $j$ denotes the quantum gate.\label{COMBO_SPIN_P}}
\end{figure*}
We performed time dependent simulations of each quantum gate that can be realized by this nanodevice taking $V_0=-0.4$mV,
which is slightly larger than for the separate nanodevices ($-0.3$mV) form previous section. The larger voltage and thus hole momentum is necessary in this case to allow the hole to pass easily through the regions between electrodes (depending on the gate we want to realize): $e_4$ and $e_6$, $e_4$ and $e_8$, $e_3$ and $e_{10}$ respectively for Pauli $X$, $Y$ and $Z$ gates. This larger initial momentum as well as the presence of additional electrodes which induce some asymmetry in the electrostatic potential distribution (lateral confinement potential) result in a ``wavy" hole trajectory which is depicted in Figs. \ref{E_POT} ($a_j$), where $j$ denotes the certain quantum gate. Fortunately, a hole trajectory that is not perfectly straight only affects the final spin state slightly. In this case the fidelity of proposed gates takes the following values: $96.8\%<\mathcal{F}_{Paili X}<99.1\%$, $98.5\%<\mathcal{F}_{Paili Y}<99.6\%$, $99.7\%<\mathcal{F}_{Paili Z}<99.9\%$, $99.1\%<\mathcal{F}_{U_s}<99.9\%$. We plot the time evolution of average HH pseudospin components $s_x(t)$, $s_y(t)$, $s_z(t)$ in Figs. \ref{COMBO_SPIN_P}(a)-(d) and the occupation probability of the different hole spin basis states: $P_{|HH\uparrow\rangle}(t), P_{|HH\downarrow\rangle}(t), P_{|LH\uparrow\rangle}(t), P_{|LH\downarrow\rangle}(t)$ can be found in Figs. \ref{COMBO_SPIN_P} (a')-(d') for each quantum gate cycle(Pauli $X$, $Y$, $Z$ and $U_S$) realized by the proposed combo nanodevice.
Application of the gate voltages as well as its geometry define the path which is passed by the hole. The area under a positively charged electrode forms a barrier for the moving hole while a negatively charged electrode forms a potential well within which the hole can be transported. In order to illustrate how the gate voltage influences the hole trajectory we plot the electrostatic potential distribution $\phi_0(x,y,z_0)$ in the quantum well region which comes from the presence of the electrodes and gate voltages applied to them at 4 crucial moments $t<t_0$, $t_{start}<t<t_{change}$, $t_{change}<t<t_{stop}<$, $t_{stop}<t$ for each (Pauli $X$, $Y$, $Z$ and $U_S$) gate cycle. The electrostatic potential $\phi_0(x,y,z_0)$ is the solution of Laplace equation in the quantum well ($z_0$) region
\begin{equation}
    \nabla^2\phi_0(x,y,z)=0
\end{equation}
with boundary conditions determined by the presence of the electrodes $\phi_0(x,y,z_{electrodes})= V_{1-11}$. We have plotted $\phi_0(x,y,z_0)$ on Fig. \ref{E_POT}.

In the presence of the hole there is an additional dip in the electrostatic distribution localized in the center of the hole wave packet and as the hole moves this dip follows the hole (self trapping mechanism). The total potential which is felt by the hole in the quantum well region was defined in Sec. II as $\Phi(x, y, z_0,t)$.

For the presented electric control scheme $t_{change}$ corresponds to a different moment of time for each quantum logic gate. In case of the Pauli $X$ gate it is reasonable to change the voltage when the hole is in the area between the $e_6$ and $e_7$ electrode and it is done in the numerical simulation at $t_{change} \approx 35$ps. For the Pauli $Y$ gate process it is convenient to choose $t_{change}\approx 35$ps, when the hole is between the electrodes $e_8$ and $e_{9}$. When the nanodevice realizes the Pauli $Z$ or $U_{S}$ ($U_{S}^{-1}$) gate, the voltage is changed at $t_{change}\approx 15$ps when the hole is under electrode $e_4$ (it is in the middle of $e_3$ at $t_{change}\approx 10$ps) just after the reflection from the first corner. The hole is stopped at $t_{stop}\approx 60$ps, $t_{stop}\approx 60$ps, $t_{stop}\approx 63$ps, $t_{stop}\approx 44$ps for Pauli $X$, $Y$, $Z$ and $U_S$ gates, respectively.

The fact that in the proposed device the quantum operations are controlled only by the weak constant voltages applied to locally defined electrodes allows for the realization of a scalable architecture. On Fig. \ref{COMBO_SCAL}(b) we plot the systems of electrodes for a scalable system of HH qubits on which each of the proposed gates can be applied in an individual, selective manner.

Furthermore, the proposed device is suitable for coherent transport of a hole wave packet and thus allows for transferring quantum information between different locations within the nanodevice. Thanks to this property two qubit gates can be realized by transporting a hole from one induced quantum dot to another one so that the two holes can occupy the same region (the hole wave functions can overlap), for example under electrode $e_1$, for a certain time $t$. Thanks to the exchange interaction their spins can swap according to the Heisenberg exchange Hamiltonian $H_s(t)=J(t)\vec{S_1}\cdot\vec{S_2}$ similar as two electron qubit gates are realized in two electron double quantum dots. More details about two electron and hole soliton dynamics as well as two qubit gate implementation in induced quantum dots and wires will be published in a forthcoming paper.

\section{Summary}
In conclusion, we proposed a set of nanodevices which can act as single quantum logic gates (Pauli $X$, $Y$, $Z$ and $U_S$) and a combo nanodevice which is capable to perform any of the Pauli $X$, $Y$, $Z$ and $U_S$ gate operations in an arbitrary sequence on a HH spin qubit. Quantum logic operations can be realized all electrically and ultrafast, i.e. within $70$ps.
The proposed devices are based on induced quantum dots and wires which allow for transporting the hole in the form of a stable soliton-like wave packet, while the hole trajectory is determined by the geometry and voltages applied to the top electrodes. The motion of the hole along specially designed paths in the presence of the Dresselhaus spin orbit field is equivalent to the sequential application of static magnetic fields which rotate the HH spin qubit. This control method allows to avoid the application of real magnetic fields which, because of the very small hole in plane g factor, have to be of the order of several Teslas, which is still experimentally challenging to achieve.

 Since quantum gates are controlled only by low static electric fields generated by the local top electrodes, our proposal can be extended to a larger number of qubits stored in the quantum register as in Fig. \ref{COMBO_SCAL}(b) where each qubit can be manipulated individually. Therefore, a scalable architecture can be realized. Furthermore, the proposed device is suitable for coherent transport of a hole wave packet, and thus allows for transferring quantum information between different locations in the nanodevice which gives perspective to couple long distant HH spin qubits and realize two qubit quantum gates in this proposed scalable system.

{\it Acknowledgements.}
This work was supported by the Polish National Science Center (Grant No. DEC-2011/03/N/ST3/02963), as well as by the ``Krakow Interdisciplinary PhD-Project in Nanoscience and Advanced Nanostructures" operated within the Foundation for Polish Science MPD Programme, co-financed by the European Regional Development Fund. This research was supported in part by PL-Grid Infrastructure.


\begin{thebibliography}{00}
\bibitem{Dq} The lateral size of CdTe induced quantum dot is about $l \approx 50$nm which is greater than quantum well height $d=10$nm in the z-direction. Cubic Dresselhaus terms can be thus neglected since its contribution is about $(d/l)^2\sim 0.04$ smaller than those of linear DSOI terms.
\bibitem{op} The operators are accurate when the hole is purely in HH states and when the hole is moving exactly straight with constant velocity along the quantum wire placed in $x$ or $y$ direction. Simulation shows that hole trajectory is not exactly straight especially after reflection from cut corner electrodes thus the operators have only approximated character. The operators are used to deduce the topology of metal gates which covers certain quantum gate.
\end{thebibliography}
\end{document}